\documentstyle[prd,aps,preprint]{revtex}
\tightenlines
\input epsf.tex
\def\DESepsf(#1 width #2){\epsfxsize=#2 \epsfbox{#1}}

\begin{document}

\draft
\preprint{\vbox{
\hbox{OSU-HEP-99-03}\hbox{CTP-TAMU-22-99}
\hbox{UMD-PP-99-111} }}
\title{Seesaw--constrained MSSM, Solution to the SUSY CP Problem and
a Supersymmetric Explanation of $\epsilon^{\prime}/\epsilon$ }

\author{ K. S. Babu$^1$, B. Dutta$^2$ and R. N. Mohapatra$^3$ }

\address{$^1$Department of Physics, Oklahoma State University,  Stillwater, OK
74078;\\
$^2$Center for Theoretical Physics, Department of Physics, Texas A \& M
University, College Station, TX  77843;\\
$^3$Department of Physics, University of Maryland, College Park, MD, 20742, USA.}
\date{May, 1999}
\maketitle
\begin{abstract}
We analyze CP violation in the supersymmetric standard model (MSSM) embedded
minimally into a left--right symmetric gauge structure with the seesaw mechanism
for neutrino masses.  With the plausible assumption of universal scalar masses
it is shown that CP violation in the hadron sector of the MSSM is described by
a single phase residing in the supersymmetry breaking Lagrangian.  This improves the CP
properties of the MSSM by providing a natural solution to the SUSY CP problem.
Furthermore, $\epsilon'/\epsilon$ vanishes in this model above the seesaw
scale; extrapolation to the weak scale then leads to a prediction in agreement
with the NA31 and the recent KTeV observations.  The electric dipole moment
of the neutron is naturally suppressed, we estimate its magnitude to be about
$4 \times 10^{-29}$ ecm.  Additional predictions include a tightly constrained
super particle spectrum and vanishingly small CP asymmetries in the $B$ meson system.
\end{abstract}
\vskip0.5in

Recent evidences for neutrino oscillations imply that the  standard model must
be extended to accommodate small neutrino masses. An elegant model that provides
an explanation of the small neutrino masses via the seesaw
mechanism\cite{see-saw}
is the left-right symmetric model of weak interaction.
Two seemingly unrelated puzzles of the standard model, viz., the
stability of the Higgs--boson mass and the question of the origin of the electroweak
symmetry breaking, seem to require its supersymmetric extension -- the
MSSM\cite{kane} -- for a proper resolution. It is well known that MSSM by itself is
plagued with a number of problems which were not present in the standard model.
In this Letter, we will be concerned with one such problem, viz., the emergence of a
plethora of arbitrary CP phases that require severe fine--tuning to explain the
smallness of observed CP violation in the Kaon system as well as the non--observation of
the electric dipole moment of the neutron\cite{susycp,nedm,aln}.
This is known as the SUSY CP
problem.  A promising approach towards its resolution would be to realize MSSM
as a low energy limit of a theory where this as well as other problems of MSSM
are cured. Here we explore supersymmetric left-right model with the seesaw
mechanism for neutrino masses as a candidate theory.

It has been pointed out that in the supersymmetric version of the
left-right model (SUSYLR)\cite{susylr,pheno} which embodies the seesaw
mechanism (and therefore explains the small neutrino masses) the
constraints of parity invariance provide a simple resolution of the SUSY
CP problem \cite{rasin}. Neutrino masses as suggested by SuperKamiokande
require that the scale above which the SUSYLR model manifests itself is at least
$10^{12}$ GeV, if the seesaw mechanism is implemented via the renormalizable
terms in the superpotential, or close to the canonical GUT scale of $10^{16}$ GeV
if one envisions the Planck  scale suppressed non--renormalizable terms as the
source of the seesaw. In either case the model could finally be embedded into
an SO(10) grand unified theory. The basic assumption of this paper is that
above the scale where the SUSYLR model is valid, the Lagrangian is
invariant under the parity transformation and below the seesaw scale the
theory is MSSM, with constraints on its parameters as required for its
embedding into the SUSYLR model.

We will focus our attention on the flavor mixing and CP violation in the minimal
versions of the high scale SUSYLR (or SO(10)) model. By minimal we mean that we
must have only one multiplet that gives rise to fermion masses, i.e. one
left--right bidoublet ({\bf 10} in the case of SO(10)). Since this one multiplet
contains both the $H_u$ and $H_d$ multiplets of the MSSM, it immediately leads
to a proportionality between the up and the down Yukawa coupling
matrices\cite{bdm1,bdm2}. This is called up--down unification. The
quark mixing angles all vanish at the tree--level, but they are induced by
loop diagrams involving the exchange of supersymmetric particles.
This considerably restricts the flavor and CP violating interactions in the model and
makes it very predictive. Note that we have not assumed the existence of any
extraneous discrete symmetries. This is one of our key starting points.

There is a second set of constraints on the model which follows from the
existence of parity invariance of the Lagrangian prior to symmetry
breaking  \cite{rasin}. Above the seesaw scale they imply that the
familiar $\mu$ and
$B$ parameters as well as the gluino mass terms are real. Furthermore, the Yukawa
coupling matrices as well as the SUSY breaking $A$ matrices (the trilinear
terms that involves the squarks) are Hermitean. In
fact, left-right symmetry implies that there is only one
$A$ matrix in the squark sector, which evolves to the two $A_{u,d}$ matrices of
low energy MSSM. It was noted in Ref.\cite{rasin} that these parity--implied
constraints solve the SUSY CP problem in the sense that the electric dipole
moment of the neutron is comfortably consistent with the present experimental
limits\cite{ramsey} without any need for fine--tuning. We thus see that
the combination of up--down unification and the constraints of parity invariance
considerably restricts flavor structure of both the quark and the squark sector
of the model\cite{bdm1,bdm2} and one might therefore expect that in addition to
solving the SUSY CP problem, there are predictions by which the model can be
tested.

In this paper we supplement the seesaw--constrained MSSM just described
with the plausible assumption of universal scalar masses.  We shall keep
the trilinear scalar $A$ terms arbitrary, subject of course to left--right symmetric
constraints. In this theory, there is only
one CP phase residing in the $A$ term that
characterizes all CP violating phenomena in the quark sector.
It is thus on par with the conventional CKM model as far as the
number of CP violating phases is concerned.

At tree level, the up and down Yukawa matrices in the model are proportional.
This results in the vanishing of the CKM angles at tree level.  Non--zero quark mixings
arise only from one--loop corrections involving the elements of the $A$ matrices.
There is no other source of flavor mixing in the model.
The $A_{ij}$ are thereby fixed to a narrow range, resulting in a very predictive
model. The value of the single CP phase is fixed by the requirement that the model
reproduce the observed value of $\epsilon_K$, the indirect CP violating
measure in  the Kaon system.
The resulting low energy theory is the MSSM, but without the SUSY CP problem and
with its parameters restricted to a very narrow range.

An interesting prediction of the model is that, owing to the
constraints of parity invariance,
$\epsilon'/\epsilon$ is vanishingly small
above the seesaw scale $v_R$. However, as the theory is RGE evolved to the weak
scale, manifest parity invariance disappears and a non--negligible value
for $\epsilon'/\epsilon$ emerges. We calculate this value and
find it to be in good agreement with the recent KTeV\cite{ktev}
and previous NA31\cite{na31} results.

A similar suppression also occurs for the electric dipole moment (edm) of
the neutron and we find that its value at the weak scale is $\sim 4\times 10^{-29}$ 
ecm. This and the fact that
up--down unification restricts the
allowed parameter space of the MSSM considerably provide
tests of the model.

Let us start by giving a brief derivation of the up--down unification relation in
the SUSYLR models. As is well known, the gauge group for this model is
$SU(2)_L\times SU(2)_R\times U(1)_{B-L}$ with the standard assignment where
$Q,Q^c$ denote left--handed and right--handed quark doublets and $\Phi$ denotes the
(2,2,0) Higgs bi--doublet. The $SU(2)_R\times U(1)_{B-L}$ symmetry could either
be broken by $B-L=2$ triplets -- the left--handed triplet $\Delta$ and the
right--handed triplet
$\Delta^c$ (accompanied by $\bar{\Delta}$ and $\bar{\Delta^c}$ fields, their
conjugates to cancel  anomalies) or by $B-L=1$ doublets $\chi $ (left) and
$\chi^c$ (right) along with $\bar{\chi}$ and $\bar{\chi^c}$.
Let us write
down the gauge invariant matter part of the superpotential involving these
fields:
\begin{eqnarray}
W & = &
{\bf Y}_q Q^T \tau_2 \Phi \tau_2 Q^c +                                   {\bf
Y}_l L^T \tau_2 \Phi \tau_2 L^c
\nonumber\\
  & +  & i ( {\bf f} L^T \tau_2 \Delta L + {\bf f}_c
{L^c}^T \tau_2 \Delta^c L^c)~.
\label{sup}
\end{eqnarray}  Below the $v_R$ scale, the $H_{u,d}$ contained in the bi--doublet
field will emerge as the MSSM doublets, but in general in arbitrary combinations
with other doublet fields in the model. The single coupling matrix $Y_q$
therefore describes the flavor mixing in the MSSM in both the up and the down
sectors leading to the relations
\begin{eqnarray} Y_u~=\gamma Y_d;
~~~~~~~~~~~~~~~Y_{\ell}~=~\gamma~Y_{\nu^D}\label{yuk}
\end{eqnarray} which we call up--down unification.
The parameter $\gamma$ is unity if the multiplets $H_u$ and $H_d$
of MSSM are contained entirely in $\Phi(2,2,0)$, but $\gamma$ can
differ from one if additional doublets contribute to $H_u$ and
$H_d$. At first sight the first of the relations in Eq. (2) might appear
phenomenologically disastrous since it leads to vanishing quark
mixings and unacceptable quark mass relations. We showed in
Ref.\cite{bdm1} that after including the one--loop corrections involving the
exchange of supersymmetric particles to the relations in Eq. (1),
there exists a large range of parameters (though not
the entire range in the MSSM) where correct quark mixings as well
as masses can be obtained. In Ref.\cite{bdm1}, we explored the
parameter space that allowed for arbitrary bilinear squark masses
and mixings as well as arbitrary form for the supersymmetry
breaking trilinear  $A$ matrix. We focused on a class of
solutions for large $\tan\beta$ $\sim 35-40$, corresponding to
$\gamma$ =1, where all quark masses mixings and CP violating
phenomena could be explained. The magnitude of $\tan \beta$ can be
reduced for $\gamma \ge 1$. In this paper, we focus on a predictive
scenario where all flavor mixing arises from the trilinear $A$ terms.
We use the small
tan$\beta$ scenario ($\gamma  \gg 1$) to explain the observed CP violation while
satisfying the FCNC constraints from the $K$ meson system.

The model becomes much more predictive
once the assumption of universal scalar masses is
imposed on the theory. This implies that at the Planck scale, the only phase of
the SUSYLR theory resides in one of the three off-diagonal entries of the SUSY
breaking trilinear coupling matrix $A$. To see how this comes about, note that
parity symmetry of the Lagrangian imposed above the $v_R$ scale implies that
$Arg (B) = Arg (\mu) =0$ and $Y_{q,l}= Y^{\dagger}_{q,l}$ and
$A_{q,l}=A^{\dagger}_{q,l}$. A Hermitean
$3\times 3$ matrix has only three independent phases. However, above the
$v_R$ scale, quark masses can be diagonalized and made real (this is because
there is only one matrix ${\bf Y}_q$ in Eq. (1).)  It therefore follows that
redefining the phases of any two quark superfields (in both the right and the left
sector) we can make two of the three off diagonal
$A_{ij}$'s real. Thus we are left with only one phase in the theory.  Thus the tree
level MSSM parameters above the $v_R$ scale can be summarized as follows (we
only discuss the quark-squark sector):
\begin{eqnarray} M^{(0)}_u &=& \gamma \tan\beta M^{(0)}_d \\ \nonumber
M^2_{\tilde{Q}}&=& M^2_{\tilde{u^c}}=  M^2_{\tilde{d^c}}=M^2_{H_{u,d}}= m^2_0 {\bf
I};\\
\nonumber M_{\tilde{W}}&=&M_{\tilde{B}}=M_{\tilde{g}}=M_{1/2} \\
 A&=&\left(\begin{array}{ccc}  A_{11} & A_{12} & A_{13} e^{i\delta^p_{13}}\\ A_{12}
& A_{22} & A_{23} \\ A_{13} e^{i\delta^p_{13}} & A_{23} & A_{33} \end{array}
\right)~.\label{A}
\end{eqnarray} The parameter $\gamma$ is related to the mixing between the
$SU(2)_L$ doublets in $\Phi$ and those in other multiplets in the theory such as
$\chi$ in the case of the non--renormalizable seesaw model or those in $(2,2,\pm
2)$ multiplets which may be included in the renormalizable seesaw
model\footnote{The advantage of including the $(2,2,\pm 2)$
multiplets instead of the $\chi$-type doublets is that the former maintain
the property of automatic R-parity conservation.}. Note that all $A_{ij}$
in Eq. (4) are real and further, we could have chosen to place the phase 
$\delta^p_{13}$ at
any off diagonal entry of $A$. However the final results are independent
of this choice.

The first task before us is to compute the one--loop corrections to the quark
mass matrices both in the up and the down sector and obtain the desired
masses and mixings. The one--loop expressions for the mass corrections to the down
type quarks  are given in the Refs.\cite{bdm1,bdm2,raby}. The up sector
corrections can be obtained by replacing
($A_d$, $\lambda_d$, $v_d$) by  ($A_u$, $\lambda_u$, $v_u$).
In the down sector, there are three types of
flavor contributions: $M_d = v_d [Y_q(1+c_1\tan\beta)
+c_2 (A_d/M_{SUSY})
+ c_3 \delta_{33}]$.  Here the $c_i$ are dimensionless loop factors.
The $c_1$ and $c_2$ terms arise from the gluino
graph, the $c_3$ term which contributes significantly only to the $b$--quark mass
is from the chargino graph.
$M_u$ is given by$M_u = v_u[Y_q(1+ c_1/\tan\beta +c_2 (A_u/M_{SUSY})]$.  Clearly
there is a mismatch between $M_u$ and $M_d$, which implies violation of
proportionality and non--zero CKM angles.

Although the scalar masses are assumed to be universal at the Planck scale, the
non--diagonal nature of the $A$ matrix will induce
via the RGE  off--diagonal elements in the
up and down squark mass matrices. Since our calculation is going to be done at the
SUSY scale of a few hundred GeV, we must extrapolate the parameters  down
from the Planck scale via the $v_R$ scale down to $M_{SUSY}$. The RGE's
for extrapolation below $v_R$ are those of MSSM and are well
known\cite{martin}.
Between $v_R\leq \mu \leq M_{P\ell}$, we use the RGE corresponding to
the SUSYLR model. We keep only the one--loop terms. It turns out that
the main effect of running from the Planck scale to $v_R$ in the squark sector
is to split the third generation squarks slightly from the first two
generations due to the large third generation
Yukawa coupling.  This effect is further amplified via the RGE
in the process of running from $v_R$ to
the SUSY scale.
We work in a basis where the tree--level Yukawa coupling matrix is
diagonal. As a result, Yukawa matrices at $M_{SUSY}$ also will be diagonal. However,
the superpartner masses which start at the Planck scale as diagonal matrices
acquire off-diagonal terms both at the $v_R$ scale and at $M_{SUSY}$. At
$M_{SUSY}$, the $\tilde{Q}$, $\tilde{u^c}$ and $\tilde{d^c}$ masses are no longer
equal. The off--diagonal entries of these matrices become complex. Similarly, the
$A$-matrix which was Hermitean at the $v_R$ scale also loses its hermiticity.
Using these extrapolated quantities, we compute the one--loop corrections to the up
and down quark mass matrices and diagonalize them to fit the known quark mixings
and masses. We find a range of input parameters at $M_{P\ell}$ which leads to the
correct quark masses and mixings. We then make sure that the chosen values of
the $A_{ij}$'s  do not lead to excessive flavor changing neutral current
effects. We have used the constraints quoted in Ref.\cite{ciu,mas}.

Before presenting an explicit example of the parameter choice and the
corresponding mass predictions, let us recall one of the main results of
 Ref.\cite{bdm1}. We showed that, to get the correct mixing angles, we
need to have
$\delta_{12,LR}\simeq 4.4\times 10^{-3}- 6.2\times 10^{-3}$,
$\delta_{23,LR}\simeq (.84-1.8)\times 10^{-2}$, where the range comes from
varying the parameter $x\equiv M^2_{\tilde{g}}/m^2_{\tilde{q}}$.
$\delta_{12,LR}$ is determined from the Cabibbo angle, $\delta_{23,LR}$ from
$V_{cb}$.
(Here $\delta_{ij}$ are the flavor violating squark mixing parameters.)
On the other
hand, from Ref.\cite{ciu}, we note that the upper bound on
$Im(\delta_{12,LR})\leq (2- 4)\times 10^{-4}$. It is then clear from this that
the phase $\delta^p_{13}$ in the $A$ matrix should be of order $10^{-2}$. Our
detailed numerical analysis also seems to generate fits to all parameters
only for a phase of this order of magnitude. Furthermore, we find that if
we scale the squark masses, the $A$ matrix and $M_{\tilde{g}}$ by a common
factor $k$, the quark mixings remain unchanged. This might lead one to suspect
that the SUSY breaking parameter range is not limited in the theory. But
since $\epsilon_K$ in our model  is coming entirely from the SUSY box
graph and it scales like $m^{-2}_{\tilde{q}}$, the squarks cannot be too heavy.
There is a scaling relation between
the CP phase and the SUSY breaking parameters. The $\epsilon'/\epsilon$ however
scales differently. As a result, we are forced to a narrow range of
the SUSY breaking masses.

To present a concrete example, we consider a case with $\tan\beta =3$,
$m_0= 80$ GeV, and $M_{1/2}=180$ GeV.  For the trilinear $A$ matrix we choose:
$(A_{11}, A_{12}, A_{13}, A_{22} A_{23}, A_{33})=(
1.2, 1.8, -2.2, -12, 17, 50)~ GeV$ and $\delta^p_{13} = 0.02$.
We determine $\mu$ from the radiative breaking of the
electroweak
symmetry.  Its magnitude in the above
parameter space  is  $\mu \simeq 290$ GeV. We choose the sign of $\mu$ as
preferred by
$b\rightarrow s\gamma$ decay.  For the quark masses and mixings, we find:
\begin{eqnarray} V_{us}&\simeq& -0.21,\,
V_{cb}\simeq 0.035,\, V_{ub}\simeq -0.0033,\,V_{td}\simeq -0.012,\,
{\rm and}\,J\simeq\sim 7\times 10^{-7}.\\
\nonumber M_d &=& ( -0.0042, -0.059, 2.62)~{\rm GeV}, ~M_{u}=(-0.0024, 0.61,
162)~{\rm GeV}.
\end{eqnarray}
The top mass (pole) is  172 GeV.  The other masses at their respective mass scales
(or at 1 GeV for $u,d,s$) can be  obtained by mulityplying with the following
QCD correction factors $\eta$:
$\eta_b$=1.59,
$\eta_c$=2.1, $\eta_s$=2.4, $\eta_d$=2.4 and $\eta_u$=2.4.  The fit for the
quark masses and mixings is quite satisfactory.

The squark mass matrices are   $6\times 6$  with   the $3\times 3$ submatrices
denoted by $M^2_{LL}$,$M^2_{RR}$ and $M^2_{LR}$.  These submatrices need to be rotated by
the  matrices  which are used to diagonalize the up and the down
quark mass matrices.  If the left--handed and the right--handed rotations for the quark
masses are  given by $U_{l,i}$ and $U_{r,i}$ where i's can be u or d, then
we write the rotated   down type squark submatrices (for example) as:
$U_{l,d}M^2_{LL}U_{l,d}^{\dag}$,
$U_{r,d}M^2_{RR}U_{r,d}^{\dag}$ and $U_{l,d}M^2_{LR}U_{r,d}^{\dag}$. We
present only the rotated
down squark mass matrices in this case since they are the only ones that enter
in the discussion of CP violation in K-decays:
\begin{eqnarray} M^2_{{LL}}&=&\left(\begin{array}{ccc} 210442 & -1.8-198 i  &
-3.5-398 i \\ -1.8+198 i & 209509 & -1594-0.86 i \\ -3.5+398 i  &
-1594+0.86 i& 174085 \end{array}\right); \\
\nonumber M^2_{{RR}}&=& \left(\begin{array}{ccc} 193030&-2.4 -263.6i  &
-5.7-651 i \\-2.4+ 263.6i& 191835 & -2583-1.4 i \\
-5.7+651 i&  -2583+1.4 i& 187244\end{array}\right); \\
\nonumber M^2_{LR} &=&\left(\begin{array}{ccc} 180+7.2\times 10^{-5}i & 0.23
+ 86.2 i & 8.85+979.8 i \\ 0.23-86.3 i & 2430+ 0.0062 i &
3060+1.72 i\\
 7.9-874.8 i & 2718.3-1.52 i & 3885.6 + 5.0\times 10^{-4} i
\end{array}\right)~.
\end{eqnarray} We need these matrices to calculate  flavor changing processes.
We find that all the flavor changing constraints arising from the SUSY exchange
are consistent with the bounds obtained in
\cite{ciu}. The six
down type squark masses in this example are given by: $(459, 458,\, 440,\,
439,\, 433\,{\rm and}\, 415)$ GeV and $M_{\tilde g} = 501$ GeV.
The parameter space of the model is quite constrained, the example above and
its overall rescaling (discussed later) are the only solution we have found.

Let us now turn to CP violation in this model. Note that as mentioned before, the
value of $\epsilon_K$ is used to determine the input phase of the theory. It is
clear from the CKM mixing matrix of the above example that the
rephasing invariant J-parameter is of order
$\sim 7\times 10^{-7}$.  If the CKM phase is to explain $\epsilon_K$,
the value needed is  $ J \sim 2\times 10^{-5}$. Thus $\epsilon_K$ has a purely
supersymmetric origin here.   The dominant contribution is from gluino box graph
involving the $LL$ and $RR$ terms in the squark mass matrix.
These $LL$ and $RR$ terms also have their origin in the off--diagonal $A$
terms through the RGE. All parameters in our model are then essentially fixed.  In
fact we have tried to vary them to see the effect. What we find is that fitting
the quark mixings essentially implies that we must vary $m_0$, $m_{1/2}$ and $A$
by a common factor $k$ relative to the example just given. The value of
$\epsilon_K$ is sensitive to $k$ since the CKM contribution to the real part of
$\Delta m_K$ is insensitive to it whereas the imaginary part of the matrix element which
receives its dominant contribution from the supersymmetric box graphs scales
like $m^{-2}_{\tilde{q}}$. Thus the only freedom allowed in our choice of squark
and gaugino masses is whatever comes from the uncertainty in the hadronic matrix
elements. Using the {12} elements of the $LL$ and $RR$ mass matrices
 of Eq. (6), we find that the
experimental bound on $\epsilon_K$ is nearly saturated. We use
Ref.\cite{mas} to find the QCD corrected bound on
$\sqrt{\left[|Im(\delta^d_{{12},{LL}}\delta^d_{{12},{RR}})|\right]}$  which is
$\sim 1.5\times 10^{-4}$ for $m_{\tilde q}\sim 460$ GeV and $x\sim 1.2$.
($\delta^d_{{12},{LL}}\equiv M^2_{{12},{LL}}/m_{\tilde q}^2$.)

Let us now turn to our prediction for $\epsilon'/\epsilon$. As usual the
dominant contributions here are from the penguin diagrams. An important point to
note is that since all parameters are now fixed, we might have a conflict with the  measured
value of  $\epsilon'/\epsilon$. The reason for this apprehension is that Masiero
and Silvastrini \cite{mas} quote an upper limit on $Im \delta_{12, LR}\simeq 10^{-5}$,
which is a factor of 10 smaller than the value required by us to fit
$\epsilon_K$. This is where exact parity symmetry comes
to the rescue. The point is that the CP-violating
$\Delta I =1/2$ penguin Hamiltonian contributed by the exchange of squarks is
proportional to the difference $\delta_{12, LR}-\delta^*_{21,LR}$, which
vanishes above the $v_R$ scale due to the constraints of parity symmetry
mentioned above. However, the RGE running makes this difference nonzero by
roughly $\sim \frac{Y^2_{33}}{16\pi^2} ln
\frac{v_R}{M_{SUSY}}$, which is at the 10\% level. This qualitative
conclusion
is borne out by our detailed numerical calculation. The diagram  for the
operator
 $\bar d_L^\alpha\sigma^{\mu\nu}t^A_{\alpha\beta}s^\beta_RG^A_{\mu\nu}$ is
formed by the squark line $\tilde d_L-\tilde b_L-\tilde s_R$ in  the
loop formed by the squark and gluino lines.  The magnitudes of the mixings are
given
 $\delta_{13,LL}(\equiv M^2_{{LL}_{13}}/\tilde m_{\tilde q}^2)$
and  $\delta_{32,LR}(\equiv M^2_{{LR}_{32}}/\tilde m_{\tilde
q}^2)$.  Similarly we have the other diagram where the operator $\bar
d_R^\alpha\sigma^{\mu\nu}t^A_{\alpha\beta}s^\beta_LG^A_{\mu\nu}$ is formed by
the squark line $\tilde d_R-\tilde b_R-\tilde s_L$. We have to
subtract one diagram from the other to calculate $\epsilon'/\epsilon$ and we
predict $\epsilon'/\epsilon \simeq 3\times 10^{-3}$ for
 the above squark mass matrix and the lattice value for the
hadronic matrix elements.

Let us now discuss the  parameter space where we can have the right
amount of CP violation in $\epsilon'/\epsilon$. One simple way is if we change
$m_0$, $M_{1/2}$ and the matrix $A$ in our example  by a common factor of $k$, then
$V_{CKM}$ and the fermion masses will
remain the same. However one needs to change the phase in order to fit
$\epsilon_K$ but $\epsilon'/\epsilon$ might go out of the experimental range.
For example if we use $k=2$, which corresponds to $m_{\tilde q}=900$ GeV, we
find that the phase $\delta^p_{13}$ is near 0.1  to fit $\epsilon_K$, however our
prediction of $\epsilon'/\epsilon$ becomes a factor 2.5 smaller than what
we had before.  If $k$ is decreased to $0.5$
(corresponds to $m_{\tilde q}=230$ GeV), then the prediction of
$\epsilon'/\epsilon$ becomes a factor of 3 larger.
The detailed predictions for the CP violating parameters
$\epsilon'/\epsilon$ along with the neutron edm $d^e_n$ for various choices of $k$ are given
in Table I. We see from the Table that $k$ somewhere between $0.7$ to $2$
is acceptable. This is the allowed spread in the squark mass parameters
and the other SUSY breaking parameters i.e. squark and gluino masses
somewhere between 300 GeV to a TeV.  Note that the ratio of the gluino mass
to the squark mass is essentially fixed, it is about 1.2 in all the fits.
This prediction could serve as a crucial test of the model.
We also find that $\tan\beta $ cannot be increased beyond about 6.
Larger $tan\beta$ would require larger value of the off diagonal elements of
$A$ (to fit $V_{cb}$, $V_{us}$ etc).  Through RGE, this would yield a SUSY
contribution to $\Delta m_K$ that is beyond the experimental limit.  Note that
the contribution to $\Delta m_K$ will grow as $(\tan\beta)^2$, so there is
really very little room for $\tan\beta \ge 6$.

Coming to the electric dipole moment of the neutron, we first note that, at
the $v_R$ scale, the flavor structure of the model is specified by the
diagonal Yukawa matrices and Hermitean $A$ matrices.  The neutron edm would
vanish in this limit, since it is given by the imaginary component of the
(11) element of the $A$ matrix.
However once we extrapolate down to
the weak scale, the situation changes and we get a nonvanishing, but quite
consistent, edm for neutron.

\newpage
Table 1. The predictions for ${\epsilon^{\prime}\over \epsilon} $ and
neutron edm  for different values of $k$. $k=1$ corresponds to $m_0=80$ GeV and
$m_{1/2}=180$GeV.

\begin{center}
 \begin{tabular}{|c|c|c|}  \hline
 $k$ & ${\epsilon^{\prime}\over \epsilon} $
 &
  Neutron edm (ecm)\\
  \hline
0.5&$1.0 \times 10^{-2}$&$5.0\times 10^{-29}$ \\\hline
1.0&$3.1\times 10^{-3}$&$5.2\times 10^{-29}$ \\\hline
1.5&$2.4\times 10^{-3}$&$6.0\times 10^{-29}$\\\hline
2.0&$1.2\times 10^{-3}$&$7.0\times 10^{-29}$\\
\hline
\end{tabular}
\end{center}

To calculate neutron edm we have
 considered 3 operators \cite{nedm}: $O_1=-{i\over2}\bar q
 \sigma_{\mu\nu}\gamma_5qF_{\mu\nu}$, $O_2=-{i\over2}\bar q
 \sigma_{\mu\nu}\gamma_5T_aqG^a_{\mu\nu}$ and $O_3={1\over6}f_{abc}
 G^a_{\mu\rho}G^{b\rho}_{\nu}G^c_{\lambda\sigma}\epsilon^{\mu\nu\lambda\sigma}$,
 where $G^x_{yz}$ is the gluon field strength and $f_{abc}$ are the Gell-mann
 coefficients.
 The effective Lagrangian with the Wilson coefficients is given by:
 $L=\sum^3_{i=1}C_i(Q)O_i(Q)$. We evaluate the $C_i$s at the weak scale and then
 we multiply by $\eta_i$ in order to evaluate them at 1.18 GeV
\cite{aln}. We use
 $\eta_1\sim$ 1.53 and $\eta_2$=$\eta_3$=3.4. Finally we use naive dimensional
 analysis\cite{mg} to calculate the quark edms ($d_q=C_1(1.18)+{e\over {4
 \pi}}C_2(1.18)+{{1.18\, e}\over {4\pi}}C_3(1.18)$) and then use the quark
models to calculate the neutron edm. We estimate $d^e_n\simeq
10^{-28}-10^{-29}$ ecm. A rough intuitive way to see this number is to
note that the dominant contribution to $d^e_n$ comes from the $A_{11}$
term which is complex and use the naive estimate from the formula
$\frac{\alpha_{s}}{4\pi}\left(\frac{m_{\tilde{g}}}{m^2_{\tilde{q}}}\right)
Im [\delta_{11,LR}]$.

Before concluding, let us comment on other models with a  supersymmetric
origin of CP violation. One of the early models of this type is that of
Ref.\cite{barr}, where all CP violation is supposed to result from the
phase of the gluino mass. Our model is different from that work since parity symmetry
makes the gluino mass real above $v_R$ and small at the weak scale.
A related class of models with approximate CP invariance \cite{nir} generically
leads to superweak CP violation which is now excluded.
The second model is the recent phenomenological analysis of Ref.\cite{mas3},
where it is noted that
a value of $\delta_{12,LR}\simeq 10^{-5}$ would produce the right value
for the $\epsilon'/\epsilon$ satisfying all other constraints.  Here
we have constructed essentially a complete theory that on RGE extrapolation leads to these parameters.
As a result, we have more specific predictions e.g. the $d^e_n$ as well
as the fact that there will be no measurable CP violation in the B-sector.
  It should be noted that while the
constraint of parity by itself does not require the Wino mass $M_2$ to
be real, when embedded into a scheme such as SO(10) where $M_2$ gets related
to $M_3$, $M_2$ becomes real and there is no additional source of CP violation.

In conclusion, we have found that the requirement that MSSM be embedded
into a supersymmetric left--right framework above a scale of
$10^{12}$ GeV  to explain the small neutrino masses observed in the
Super-Kamiokande experiment, imposes very stringent constraints on the
parameters of the MSSM. First it predicts the value of
$\epsilon'/\epsilon$ in agreement with experiment and the edm of neutron
comfortably consistent with the present upper limits. We also find the
super--partner masses to be in a very narrow range
with the gluino mass not much different from the squark masses,
which can provide a test
of the model.  The parameter $\tan\beta$ is not expected to exceed about 6.
In the CP violation sector, a test of the model is the
absence of significant CP violating effects in the B-sector.
The standard KM contribution is negligible, so is the supersymmetric contribution
(see Eq. (6)). These models
can be readily unified into SO(10) models or other unification groups that contain
the SUSY left-right model as a subgroup in them.

The work of KSB is supported by funds from the Oklahoma State University.
RNM is supported by the National Science Foundation grant No.
PHY-9802551.

\end{document}